\documentclass[12pt]{article}
\usepackage{amsfonts}
\usepackage{amsgen,amsmath,amstext,amsbsy,amsopn,amssymb,soul,subfigure}

\usepackage[pdftex]{graphicx}
\usepackage{epstopdf}
\usepackage{natbib} 
\usepackage[dvipsnames,table,dvipsnames*, svgnames*]{xcolor}
\usepackage{comment}

\definecolor{blue}{RGB}{050,050,250}
\definecolor{green}{RGB}{000,150,100}
\definecolor{purple}{RGB}{220,040,250}

\def\red{\color{red}}

 \usepackage{natbib,lscape,longtable}
 \usepackage{algorithmic}
 \usepackage[]{algorithm2e}
 \usepackage{multirow}
 
 \usepackage[]{inputenc}
 \bibpunct{(}{)}{;}{a}{,}{,}
 
 \usepackage{adjustbox}
 \usepackage{booktabs}
 \newcommand{\ignore}[1]{}
 
 \newcommand{\Exp}{\mathbb{E}}

 \newcommand{\Prob}{\mathbb{P}}

 \newcommand{\argmin}{\operatornamewithlimits{argmin}}
 
 \newcommand{\bX}{\mathbf{X}}
 \newcommand{\bZ}{\mathbf{Z}}
 \newcommand{\bC}{\mathbf{C}}
 \newcommand{\bV}{\mathbf{V}}
 \newcommand{\bM}{\mathbf{M}}
 \newcommand{\bA}{\mathbf{A}} 
  \newcommand{\bP}{\mathbf{P}} 
 \newcommand{\bSigma}{\boldsymbol{\Sigma}}
 \newcommand{\bOmega}{\boldsymbol{\Omega}}
 \newcommand{\bLambda}{\boldsymbol{\Lambda}}
 \newcommand{\btZ}{\widetilde{\mathbf{Z}}}
 
 \newcommand{\biden}{\mathbf{I}_p}
 \newcommand{\tM}{\mathbf{\widetilde{M}}}
 
 \newtheorem{theorem}{Theorem}
 \newtheorem{lemma}{Lemma}
 \newtheorem{corollary}{Corollary}
 \newtheorem{definition}{Definition}
 \newtheorem{condition}{Condition}

\begin{document}
 		\title{Regression Analysis for Microbiome Compositional Data$^1$}
 		\author{Pixu Shi, Anru Zhang and Hongzhe Li}
 		
 		\date{}
 		
 		\maketitle

\footnotetext[1]{Pixu Shi is with Department of Biostatistics and Epidemiology, University of Pennsylvania; Anru Zhang is with Department of Statistics, University of Wisconsin-Madison; Hongzhe Li is with Department of Biostatistics and Epidemiology, University of Pennsylvania. This research was supported by NIH grants CA127334 and GM097505.}
 		
 			
 			
 			
 			


\begin{abstract}
One important problem in microbiome analysis is to identify the bacterial taxa that are associated with a response, where the microbiome data are summarized as the composition of the bacterial taxa at different taxonomic levels.  This paper considers regression analysis with such compositional data as covariates. In order to satisfy the subcompositional coherence of the results, linear models with  a set of linear constraints on the regression coefficients are introduced.  Such models allow regression analysis for subcompositions and include the log-contrast model for compositional covariates as a special case. A penalized estimation procedure for estimating the regression coefficients and for selecting variables under the linear constraints is developed. A method is also  proposed to obtain de-biased estimates of the regression coefficients that are asymptotically unbiased and have a joint asymptotic multivariate normal distribution. This  provides  valid confidence intervals of the regression coefficients and can be used to obtain the $p$-values. Simulation results show the validity of the confidence intervals and smaller variances of the de-biased estimates when the linear constraints are imposed. The proposed methods are applied to  a gut microbiome data  set and  identify four  bacterial genera that are associated with the body mass index after adjusting for the total fat and caloric intakes. 

\end{abstract}

\noindent{\bf Keywords:\/} Compositional coherence; Coordinate descent method of multipliers; High dimension; Log-contrast model; Model selection; Regularization; 


\section{Introduction}

The human microbiome  includes all microorganisms in and on the human body. These microbes  play important roles in  human metabolism, nutrient intake and energy generation and thus are essential in human health. The gut microbiome has been shown to be associated with many human diseases such as obesity, diabetes and inflammatory bowel disease  \citep{Turn:Ley:Maho:Magr:Mard:Gord:obes:2006,qin2012metagenome,manichanh2012gut}.   Next generation sequencing technologies make it possible to study the microbial compositions without the need for  culturing  the bacterial species.   There are, in general, two approaches to quantify the relative abundances of bacteria in a community. One approach is based on sequencing the 16S ribosomal RNA (rRNA) gene, which is ubiquitous in all bacterial genomes. The resulting sequencing reads provide information about the  bacterial taxonomic composition. Another approach is  based on shotgun metagenomic sequencing, which sequences all the microbial genomes presented in the sample, rather than just one marker gene. Both 16S rRNA and shotgun sequencing approaches provide  bacterial taxonomic composition information  and have been widely applied to human microbiome studies, including the Human Microbiome Project (HMP) \citep{turnbaugh2007human} and the Metagenomics of the Human Intestinal Tract (MetaHIT) project \citep{qin2010human}. 

Several methods are available for quantifying the microbial relative abundances based on the sequencing data, which typically involve aligning the reads to some known database \citep{segata2012metagenomic}. Since the DNA yielding materials are different across different samples, the resulting numbers of sequencing reads vary greatly from sample to sample. In order to make the microbial abundance comparable across samples, the abundances in read counts are usually normalized to the relative abundances of all bacteria observed. This results in high-dimensional  compositional data with a unit sum. Some of the most widely used metagenomic processing softwares such as MEGAN \citep{huson2007megan} and MetaPhlAn \citep{segata2012metagenomic} only output the relative abundances of the bacterial taxa at different taxonomic levels.  

This paper considers regression analysis of microbiome compositional data, where the goal is to identify the bacterial taxa that are associated with a continuous response such as the body mass index (\textsc{bmi}). Compositional data are strictly positive and multivariate that are constrained to have a unit sum.  Such data are also referred to as mixture data \citep{Aitc:Baco:log:1984, Snee, Cornell}. Regression analysis with compositional covariates needs to account for 
the intrinsic multivariate nature and the inherent interrelated structure of such data.  For compositional data, it is impossible
to alter one proportion without altering at least one of the other proportions.  Linear log-contrast model \citep{Aitc:Baco:log:1984} has been proposed for compositional data regression where logarithmic-transformed proportions are treated as covariates in a linear regression model with the constraint of the sum of the regression coefficients being zero. \cite{LinLi} proposed a variable selection procedure for such models in high-dimensional settings and  derived the weak oracle property of the resulting estimates.  In analysis of microbiome data, it is also of biological interest to study the subcompositions of bacteria taxa within  higher taxonomic levels, such as subcompositions of species under a given genus or phylum, or subcompositions of genera within a phylum.  In subcompositional data, the  proportions of species have been calculated
relative to total proportions of the species under a given genus; that is, the values in the subcomposition have been “re-closed”
to add up to 1.  Regression analysis of such subcompositional data is also considered in this paper. 

 One of the founding principles of
compositional data analysis is that of subcompositional coherence \citep{Aitc:stat:1982}:  any compositional data analysis should be done in a way that we obtain
the same results in a subcomposition, regardless of whether we analyze only that
subcomposition or a larger composition containing other parts. This is especially relevant in high-dimensional regression analysis with compositional covariates, where the goal is to select the bacteria whose compositions are associated with the response.  Once such bacteria are identified, it is desirable to recalculate the subcomposition only within those identified. However,  these subcompositions have different values from those calculated based on a larger set of bacterial taxa.  The log-contrast model of \cite{Aitc:Baco:log:1984}  and \cite{LinLi} satisfies this principal by imposing a linear constraint on the regression coefficients.  This paper extends  this model for analysis of microbiome  subcompositions, where multiple linear constraints are imposed  in order to achieve the subcompositional coherence. 

Penalized and constrained regression, including  constrained Lasso regression,  has been studied by  \cite{james2014penalized}, where the regression coefficients are subject to a set 
of linear constraints.   A computational algorithm through reformulating the problem as an unconstrained optimization problem was proposed and  non-asymptotic error bounds of the estimates were  derived.  Different from \cite{james2014penalized}, this paper presents  an efficient computational algorithm based on the coordinate descent method of multipliers and augmented Lagrange of optimization problem.  Since the resulting estimates are often biased due to $\ell_1$ penalty imposed on the coefficients,  variance estimation  and statistical inference of the resulting estimates are difficult to derive.  In order to make the statistical inference on  the regression coefficients and to obtain the confidence intervals,  asymptoticly  unbiased estimates of the regression coefficients are first obtained  through a de-biased procedure and their joint asymptotic distribution is derived.  The proposed de-biased procedure extends that of \cite{Javanmard:2014} to take into account the linear constraints on regression coefficients.   However, due to the linear constraints on the regression coefficients, the theoretical developments are different from \cite{Javanmard:2014}. 

Section \ref{models} presents linear regression models with linear constraints for compositional covariates.  Section \ref{estimation} presents an efficient coordinate descent method of multipliers  to implement the penalized estimation of the regression coefficients under linear constraints.  Section \ref{theory} provides an algorithm to obtain de-biased estimates of the coefficients and derives their joint asymptotic distribution.  Section \ref{real}   presents results from an analysis of gut microbiome data set in order to identify the bacterial genera that are associated with \textsc{bmi}.  
Methods are evaluated in Section \ref{simulation} through simulations. 

\section{Regression Models for Compositional Data}\label{models}
\subsection{Linear log-contrast model}
Linear log-contrast model \citep{Aitc:Baco:log:1984} has been proposed for compositional data regression. Specifically, suppose an $n\times p$ matrix $\bX$ consists of $n$ samples of the composition of mixture with $p$ components, and suppose $Y$ is a response variable depending on $\bX$. The nature of composition makes each row of $\bX$ lie in a $(p-1)$-dimensional positive simplex $S^{p-1} = \{(x_1,\dots,x_p):x_j>0, j=1,\dots,p \mbox{ and } \sum_{j=1}^p x_j=1\}$. Based on this nature, \cite{Aitc:Baco:log:1984} introduced a linear log-contrast model as follows:
\begin{equation}\label{eq:model00}
Y=\bZ^p\beta_{\backslash p} + \epsilon,
\end{equation}
where $\bZ^p = \{\log(x_{ij}/x_{ip})\}$ is $n\times (p-1)$ log-ratio matrix with the $p$th component as the reference component, $\beta_{\backslash p}=(\beta_1,\dots,\beta_{p-1})$ is the regression coefficient vector, and noise $\epsilon$ is independently distributed as $N(0,\sigma^2)$. An intercept term is not included in the model, since it can be
eliminated by centering the response and predictor variables.

The selection of reference component is crucial to analysis, especially in high-dimensional settings. To avoid  choosing  an arbitrary   reference component, \cite{LinLi} reformulated model~ \eqref{eq:model00} as a regression problem with a linear constraint on the coefficients by  letting  $\beta_p = -\sum_{j=1}^{p-1} \beta_j$,

\begin{equation}\label{eq:model0}
Y=\bZ\beta + \epsilon, \quad 1_p^\top \beta=0,
\end{equation}
where $1_p=(1,\dots,1)^\top\in \mathbb{R}^p$, $\bZ = (z_1,\dots,z_p)=(\log x_{ij})\in\mathbb{R}^{n\times p}$, and $\beta=(\beta_1,\dots,\beta_p)^\top$.

\subsection{Subcompositional regression model}
In analysis of microbiome data, the relative abundances of taxa are often obtained at different taxonomic ranks, including species, genus, family, class and phylum.  It is of interest  to study whether the composition of taxa  that belong to a given taxon at a higher  rank  is associated with the response, in which case subcompositions of taxa (e.g., all the genera that belong to a given phylum) are calculated.  Suppose  $r$ taxa at a given rank are  considered with $m_g$ taxa at the lower rank that belong to taxon  $g$.   
Let 	$X_{gs}$ be the relative abundance of the $s$th taxon  that belong to the $g$th taxon at a higher rank,  for 	$g=1,\cdots, r$, $s=1,\cdots,m_g$ such that
	$$\sum_{s=1}^{m_g}{X_{gs}}=1,\mbox{for } g=1,\cdots,r.$$  Let $n\times m_g$ matrix $\bX_{g}$ represents  $n$ samples of the subcomposition  of $m_g$ taxa. 
The following model can be used to link the subcompositions to a response $Y$,
\begin{equation}\label{model2}
Y=\sum_{g=1}^{r} \bZ_{g}\beta_g + \epsilon,
\end{equation}	
where $\bZ_{g}=(Z_{g1},\dots, Z_{gm_g})=(\log X_{g1},\dots,\log X_{gm_g})\in  \mathbb{R}^{n\times m_g}$, and $\beta_g=(\beta_{g1}, \cdots, \beta_{gm_g})^\top$. To make the model subcompositional coherence,   the following $r$ linear constraints are imposed, 
	$$\mathbf{1}_{m_g}^\top \beta_g=\sum_{s=1}^{m_g} \beta_{gs}=0 \mbox{ for } g=1\cdots,r.$$
	This set of linear constraints can be written as $\bC^\top \beta=0$, where 
	$\beta=(\beta_1^\top, \cdots, \beta_r^\top)^\top$, and 
	$$\bC^\top=\left(\begin{array}{ccccccccc}
	1&\cdots&1&0&\cdots&0&0&\cdots &0\\
		0&\cdots&0&1&\cdots&1&0&\cdots &0\\
 	\vdots&\vdots&\vdots&\vdots&\cdots&\vdots&\vdots&\vdots &\vdots\\
0&\cdots&0&0&\cdots&0&1&\cdots &1\\
	\end{array}
	\right)_{r\times p}$$
	
Models~\eqref{eq:model0} and ~\eqref{model2} belong to a more general high-dimensional linear model with $r$ linear constraints on the coefficients,
\begin{equation}\label{eq:gmodel0}
Y=\bZ\beta + \epsilon, \quad \bC^\top \beta=0,
\end{equation}
where the rows of $\bZ\in \mathbb{R}^p$ are independently and identically distributed with mean zero, $\bC$ is a $p\times r$ matrix of the constraint coefficients, $\beta=(\beta_1,\dots,\beta_p)^\top$, and $\epsilon\sim N_n(0,\sigma^2 \mathbf{I})$.
Without loss of generality, $\bC=(c_1,\dots,c_r)$ is assumed to be orthonormal. In high-dimensional settings,
 $\beta$ is  assumed to be $s$-sparse, where $s=\#\{i: \beta_i\ne 0\}$ and  $s=o(\sqrt{n}/\log p)$.

This paper considers  estimation and inference of Model~\eqref{eq:gmodel0} under the general linear constraints. 
\cite{LinLi} proposed a procedure for  variable selection and estimation for Model~\eqref{eq:model0} and  derived the weak oracle property of the resulting estimates. \cite{james2014penalized} considered a more general model and provided non-asymptotic bounds on estimation errors. However,  variances of  the estimates and statistical inference are lacking. In this paper, an algorithm to perform variable selection for Model~\eqref{eq:gmodel0} based on $\ell_1$ penalized estimation is first proposed based on coordinate descent method of multipliers. An inference procedure for the penalized estimator of the regression coefficients is then introduced.   The proposed  approach parallels to that of \cite{Javanmard:2014} by first obtaining  de-biased estimates of the coefficients for high-dimensional linear model with linear constraints, $\hat{\beta}^u$, which are shown to be asymptotically Gaussian, with mean $\beta$ and covariance $\sigma^2(\widetilde{\bM}\widehat{\bSigma}\widetilde{\bM})/n$, where $\widehat{\bSigma}$ is the empirical covariance and $\widetilde{\bM}$ is determined by  solving a convex program. Based on this asymptotic result,  the corresponding confidence intervals and $p$-values are constructed and used for statistical inference.

\section{Penalized Estimation}\label{estimation}
In this following presentation, for a matrix $\bA_{m\times n}$, $||\bA||_p$ is the $\ell_p$ operator norm defined as 
$||\bA||_p=\sup_{||x||_p=1}||\bA x||_p$, where $||v||_p$ is the standard $\ell_p$ norm  of a vector $v$. In particular, 
$||\bA||_{\infty}=\max_{1\le i\le m}\sum_{j=1}^n |a_{ij}|$. We also define $|\bA|_{\infty}=\max_{i,j}|a_{ij}|$.

Consider model~\eqref{eq:gmodel0}.
Define $\bP_\bC=\bC\bC^\top$ as the projection onto the space spanned by the columns of $\bC$. Two basic regularity conditions on  $\bC$ are assumed: 
\begin{condition}\label{cond1}
$||\biden-\bP_\bC||_\infty\leq k_0$ for a constant $k_0$ that is free of $p$. 
\end{condition}
\begin{condition}\label{cond2}
The diagonal elements of $\biden - \bP_\bC$ are greater than zero.
\end{condition}
Condition 1 is equivalent to that $||c_j||_1||c_j||_{\infty}, j=1,\dots,r$ are all bounded by a constant that is free of $p$. Condition 2 means that the group of constraints do not indicate simple constraint such as $\beta_j=0$. If $(\biden-\bP_\bC)_{j,j}=0$, then the $j^{th}$ row and column of $\biden-\bP_\bC$ are all zeros, and thus $(\biden-\bP_\bC)e_j=0$, which means that $e_j$ lies in the space spanned by the columns of $\bC$. It is easy to verify that the constraint matrix $\bC$ in the log-contrast model~\eqref{eq:model0} or 
the subcompositional model~\eqref{model2} satisfies both conditions. For example, 
in the log-contrast model~\eqref{eq:model0},  $k_0=2$ for $\bC=1_p/\sqrt{p}$  since 
$$||(\biden-1_p1_p^\top/p)a||_{\infty}=||a-\dfrac{1}{p}\sum_{j=1}^pa_j1||_{\infty}\leq ||a||_{\infty}+|\dfrac{1}{p}\sum_{j=1}^pa_j|\leq 2||a||_{\infty}.$$

Define $\btZ = \bZ (\biden-\bP_\bC)$. Since $\bP_\bC\beta=0$, model~\eqref{eq:gmodel0} can be rewritten as
\begin{equation}\label{eq:gmodel}
Y=\btZ\beta + \epsilon, \quad \bC^\top \beta=0.
\end{equation}
The regression coefficients can be estimated using $\ell_1$ penalized estimation with linear constraints,
\begin{align}
\widehat{\beta}^n=\argmin_{\beta}\left(\dfrac{1}{2n}||Y-\btZ\beta||_2^2+\lambda||\beta||_1\right) \mbox{ subject to } \bC^\top \beta=0,\label{eq:l1opt}
\end{align}
where $\lambda$ is a tuning parameter. 

\ignore{Denote $\bC^{\perp}$ as the matrix with columns as a set of basis of the perpendicular space of the column space of $\bC$. If we reformulate the optimization problem~\eqref{eq:l1opt} into the optimization problem below, then it can be solved by generalized lasso (Tibshirani 2011),
\begin{align*}
\widehat{\beta}^n_c&=\argmin_{\beta_c}\left(\dfrac{1}{2n}||y-\btZ_c\beta_c||_2^2+\lambda||\bC^{\perp}\beta_c||_1\right)\\
\widehat{\beta}^n&=\bC^{\perp}\widehat{\beta}^n_c
\end{align*}
where $\btZ_c = \btZ \bC^{\perp}$, and $\beta_c = (\bC^{\perp})^\top\beta$.}

\ignore{\red The $\ell_1$ penalized estimation with linear constraints received much attention in the last decade. Many theoretical guarantees in terms of estimation loss bound has been proposed under various conditions, including restricted isometry property, restricted eigenvalue. Different computational results based on various algorithms, such as LARS, coordinate decent, has been proposed to solve the optimization efficiently. See \cite{james2012constrained,james2014penalized} and the references therein. Here,} 

A coordinate descent method of multipliers can be used to implement the constrained optimization problem~\eqref{eq:l1opt}. 
First, the augmented Lagrange of optimization problem~\eqref{eq:l1opt} \citep{Bert:cons:1996} is formed as,
$$L_{\mu}(\beta,\eta) = \dfrac{1}{2n}||y-\btZ\beta||_2^2+\lambda||\beta||_1+\eta^\top \bC^\top\beta+\dfrac{\mu}{2}||\bC^\top\beta||_2^2,$$
where $\eta\in\mathbb{R}^r$ is the Lagrange multiplier, and $\mu>0$ is a penalty parameter.  Problem~\eqref{eq:l1opt} can be solved by iterations
$$\beta^{k+1}\leftarrow \argmin_{\beta} L_{\mu}(\beta,\eta^k), \quad \eta^{k+1}\leftarrow \eta^k+\mu \bC^\top\beta^{k+1}.$$
Define $\xi=\eta/\mu$, the iterations become
\begin{align}
\beta^{k+1}&\leftarrow \argmin_{\beta}\left\{\dfrac{1}{2n}||y-\btZ\beta||_2^2+\lambda||\beta||_1 + \dfrac{\mu}{2}||\bC^\top\beta+\xi^k||_2^2\right\},\label{eq:update_beta}  \\
\xi^{k+1}&\leftarrow\xi^k+\bC^\top\beta^{k+1}.\label{eq:update_xi} 
\end{align}
The iteration of $\beta$ can be further detailed as
\begin{equation}\label{eq:update_beta_step}
\beta_j^{k+1} \leftarrow \dfrac{1}{\dfrac{||\tilde{z}_j||_2^2}{n}+\mu ||C_j||_2^2}S_{\lambda}\left[\dfrac{1}{n}\tilde{z}_j^\top(y-\sum_{i\neq j}\beta_i^{k+1}\tilde{z}_i)-\mu(\sum_{i\neq j}\beta_i^{k+1}C_i^\top C_j+C_j^\top\xi^k)\right],
\end{equation}
where $C_i, i=1,\dots,p$ are the rows of $\bC$, $\tilde{z}_i, i=1,\dots,p$ are columns of $\btZ$, and $S_{\lambda}(t)=\mbox{sgn}(t)(|t|-\lambda)_+$.
Combining~\eqref{eq:update_beta}-\eqref{eq:update_beta_step} yields the following algorithm for solving problem~\eqref{eq:l1opt}.

\begin{algorithm}[H]
	\hrulefill\caption{\it Coordinate descent method of multipliers for solving problem~\eqref{eq:l1opt}}\label{algo1}
 \textbf{Input:} $Y$, $\btZ$, and $\lambda$.\\
\textbf{Output:} $\widehat{\beta}^n$
\begin{algorithmic}[1]
\STATE
Initialize $\beta^0$ with 0 or a warm start, $\xi^0=0$, $\mu>0$ and $k=0$.
\STATE
For $j=1,\dots,p,1,\dots,p,\dots$, update $\beta_j^{k+1}$ by~\eqref{eq:update_beta_step} until convergence.
\STATE
Update $\xi^{k+1}$ by~\eqref{eq:update_xi}.
\STATE
$k\leftarrow k+1$ and repeat the two steps above until convergence.
\end{algorithmic}
	\hrulefill
\end{algorithm}

The penalty parameter $\mu$ that is needed to enforce the zero-sum constraints does not affect the
convergence of Algorithm 1 as long as $\mu>0$. It can however affect the convergence rate of the algorithm. In this paper, $\mu=1$ is taken in all the computations.

\section{A De-biased Estimator and Its Asymptotic Distribution} \label{theory}

\ignore{\subsection{Definitions}
We first introduce some definitions.

\begin{definition}
For a random variable $X$, the sub-Gaussian norm of $X$ is defined as
$$||X||_{\psi_2} = \sup_{q\geq 1}q^{-1/2} (\Exp|X|^q)^{1/q}.$$
The sub-Gaussian norm of a random vector $\mathbf{X}\in\mathbb{R}^n$ is defined as
$$||\mathbf{X}||_{\psi_2} = \sup\{||X^\top x||_{\psi_2}: x\in\mathbb{R}^n, ||x||_2=1\}.$$
\end{definition}

\begin{definition}
For any matrix $\bM$, define the upper and lower restricted isometry property (RIP) constants of order $k$ as below respectively
$$\delta_k^+(\bM)=\sup\left\{\dfrac{||\bM\alpha||_2^2}{||\alpha||_2}: \alpha \mbox{ is } k \mbox{-sparse vector}\right\},$$
$$\delta_k^-(\bM)=\inf \left\{\dfrac{||\bM\alpha||_2^2}{||\alpha||_2}: \alpha \mbox{ is } k \mbox{-sparse vector}\right\}.$$
\end{definition}

}

\subsection{A De-biased estimator}
The asymptotic distribution of $\ell_1$ regularized estimator $\widehat{\beta}^n$ is not manageable and $\widehat{\beta}^n$ is biased due to regularization.   \cite{Javanmard:2014} proposed a procedure to construct a de-biased version of the unconstrained \textsc{lasso} estimator that has a tractable asymptotic distribution, which can be used to obtain the  confidence intervals of the regression coefficients.  Similar de-biased procedures were also developed by \cite{ZhangZhang} and \cite{Buhl}.

Adapting  the de-biased procedure of \cite{Javanmard:2014},  the following algorithm can be used to  obtain  de-biased estimates of the regression coefficients, $\widehat{\beta}^u$.
\\
\begin{algorithm}[H]
	\hrulefill
\caption{\it Constructing a de-biased estimator}\label{algo2}
\textbf{Input:} $Y$, $\bZ$, $\widehat{\beta}^n$, and $\gamma$.\\
\textbf{Output:} $\widehat{\beta}^u$
\begin{algorithmic}
\STATE
Let $\widehat{\beta}^n$ be the regularized estimator from optimization problem~\eqref{eq:l1opt}.
\STATE
Set $\btZ=\bZ(\biden-\bP_\bC)$.
\STATE
Set $\widehat{\bSigma}\equiv (\btZ^\top \btZ)/n$.
\STATE
\textbf{for} $i=1,2,\dots,p$ \textbf{do}:\\
Let $m_i$ be a solution of the convex program:
\begin{equation}\label{eq:opt}
\begin{split}
\mbox{minimize }& m^\top\widehat{\bSigma}m\\
\mbox{subject to }& ||\widehat{\bSigma}m-(\biden-\bP_\bC )e_i||_{\infty}\leq \gamma.
\end{split}
\end{equation}
\textbf{end for}
\STATE
Set $\bM=(m_1,\dots,m_p)^\top$, set
\begin{equation}\label{eq:tildeM}
\tM=(\biden-\bP_\bC)\bM(\biden-\bP_\bC).
\end{equation}
\STATE
Define the estimator $\widehat{\beta}^u$ as follows:
\begin{equation}\label{eq:algorithm}
\widehat{\beta}^u=\widehat{\beta}^n+\dfrac{1}{n}\tM\btZ^\top(Y-\btZ\widehat{\beta}^n).
\end{equation}
	\hrulefill\end{algorithmic}
\end{algorithm}

To solve problem (\ref{eq:opt}),   Matlab package 
\textsc{cvx} is used for specifying and solving convex programs \citep{CVX}. 
To briefly explain the logic behind this algorithm, denote $\bSigma=\Exp \btZ^\top \btZ$, and suppose that $\bSigma=\bV\bLambda \bV^\top$ is the eigenvalue/eigenvector  decomposition of $\bSigma$, where $\bLambda=diag(\lambda_1,\dots,\lambda_{p-r})$. Note that $(\bV,\bC)$ is full rank and orthonormal, and
$$\bSigma=(\bV,\bC)\left(\begin{array}{cc}
\bLambda & 0 \\
0 & 0
\end{array}\right)(\bV,\bC)^\top. $$
Define
$$\bOmega=(\bV,\bC)\left(\begin{array}{cc}
\bLambda^{-1} & 0 \\
0 & 0
\end{array}\right)(\bV,\bC)^\top, $$
then
$$\bSigma\bOmega=(\bV,\bC)\left(\begin{array}{cc}
\mathbf{I}_{p-r} & 0 \\
0 & 0
\end{array}\right)(\bV,\bC)^\top=\bV\bV^\top=\biden-\bP_\bC,$$
where $\bOmega$ is the inverse of $\bSigma$ in the perpendicular space of the column space of $\bC$. The de-biased algorithm first finds an approximation of $\bOmega$ by rows, denoted by $\widetilde{\bM}$, and then corrects the bias based on $\widetilde{\bM}$.
At the last step of this algorithm, $\widehat{\beta}^u$ is the de-biased version of $\widehat{\beta}^n$. It is easy to check that $\bC^\top \widehat{\beta}^u=0$, which is guaranteed by~\eqref{eq:tildeM}.

The feasibility of the optimization~\eqref{eq:opt} is presented in Lemma \ref{th:feasible} under the following assumptions on matrix $\btZ=(\widetilde{Z}_1,\dots,\widetilde{Z}_n)^\top$:
\begin{condition}\label{cond3}
{There exist uniform constants} $C_{\min}, C_{\max}$ such that $0< C_{\min}\leq \sigma_{\min}(\bSigma) \leq \sigma_{\max}(\bSigma)\leq C_{\max} < \infty $, where $\sigma_{\max}(\bA) (\sigma_{\min}(\bA) )$  is the largest (smallest) non-zero eigenvalue of matrix $\bA$.
\end{condition}
\begin{condition}\label{cond4}
{There exists a uniform constant $\kappa\in (0,\infty)$} such that the rows of $\btZ\bOmega^{1/2}$ are sub-Gaussian with $||\bOmega^{1/2}\widetilde{Z}_1||_{\psi_2}\leq \kappa$, where  the sub-Gaussian norm of a random vector ${Z}\in\mathbb{R}^n$ is defined as
$$||{Z}||_{\psi_2} = \sup\{||{Z}^\top x||_{\psi_2}: x\in\mathbb{R}^n, ||x||_2=1\},$$
with $||X||_{\psi_2}$ defined as $||X||_{\psi_2} = \sup_{q\geq 1}q^{-1/2} (\Exp|X|^q)^{1/q}$ for a random variable $X$.
\end{condition}
These two conditions are imposed on $\btZ = \bZ (\biden-\bP_\bC)$, not on the original log-ratio matrix $\bZ$.  For the subcompositional model~\eqref{model2}, it is easy to see that $\btZ$  is  the  matrix of the centered log-ratio (CLR)  transformation of the original taxonomic composition \citep{Aitc:stat:1982}, where 
$$\btZ_{gs}=\log \frac{X_{gs}}{\sqrt[m_g]{\prod_{s=1}^{m_g}X_{gs}}}.$$
CLR has been shown to be effective in transforming compositional data  to approximately  multivariate normal in many real compositional and microbiome data \citep{Aitc:stat:1982, Bonneau}.  Conditions \ref{cond3} and \ref{cond4} are therefore reasonable assumptions in our setting. 

The following Lemma  shows that if $\gamma=c\sqrt{\log p/n}$ in Algorithm 2 is properly chosen, $\bOmega$ is in the feasible set of the optimization problem~\eqref{eq:opt}  with a large probability.

\begin{lemma}\label{th:feasible}
Let $\widehat{\bSigma}\equiv (\btZ^\top \btZ)/n$ be the empirical covariance. For any constant $c>0$, the following holds true,
$$\Prob \left\{\left|\bOmega\widehat{\bSigma}-(\biden-\bP_\bC)\right|_{\infty}\geq c\sqrt{\dfrac{\log p}{n}}\right\} \leq 2p^{-c''},$$
where 
 $c''=(c^2C_{\min})/(24e^2\kappa^4C_{\max})-2$.
\end{lemma}

\subsection{Asymptotic distribution and inference}

To obtain the asymptotic distribution of the de-biased estimator $\widehat{\beta}^u$,  an additional assumption on $\btZ$ is required. 
\begin{condition}\label{cond5} The inequality 
$(3\tau-1)\delta^{-}_{2s}(\btZ/\sqrt{n})-(\tau+1)\delta^{+}_{2s}(\btZ/\sqrt{n})\geq 4\tau \phi_0$ holds for a constant $\phi_0>0$,
where for any matrix $\bA\in\mathbb{R}^{n\times m}$, $\delta_k^+(\bA)$ and $\delta_k^-(\bA)$ are the upper and lower restricted isometry property (RIP) constants of order $k$ defined as
$$\delta_k^+(\bA)=\sup\left\{\dfrac{||\bA\alpha||_2^2}{||\alpha||_2^2}: \alpha\in\mathbb{R}^m \mbox{ is } k \mbox{-sparse vector}\right\},$$
$$\delta_k^-(\bA)=\inf \left\{\dfrac{||\bA\alpha||_2^2}{||\alpha||_2^2}: \alpha\in\mathbb{R}^m \mbox{ is } k \mbox{-sparse vector}\right\}.$$
\end{condition}
Condition \ref{cond5} means that $\delta^{-}_{2s}(\btZ/\sqrt{n})$ and $\delta^{+}_{2s}(\btZ/\sqrt{n})$ should be close, that is, any $2s$ columns of the CLR transformed compositional data matrix $\btZ/\sqrt{n}$ should be close to orthonormal.

The following theorem gives the asymptotic distribution of the de-biased estimates of the regression coefficients. 
\begin{theorem}\label{th:main}
Consider the linear model~\eqref{eq:gmodel} with $\beta$ as an $s$-sparse vector, and let $\widehat{\beta}^u$ be defined as in equation~\eqref{eq:algorithm} in Algorithm \ref{algo2}. Then,
$$\sqrt{n}(\widehat{\beta}^u-\beta)=B+\Delta,\quad
 B|\bZ\sim N(0,\sigma^2\tM\widehat{\bSigma}\tM^\top),\quad
 \Delta=\sqrt{n}(\tM\widehat{\bSigma}-(\biden-\bP_\bC))(\beta-\widehat{\beta}^n).$$
Further, assume the Conditions~\eqref{cond1}-\eqref{cond5} hold.
Then setting $\lambda=r\tilde{c}\sigma\sqrt{(\log p)/n}$ in optimization problem~\eqref{eq:l1opt} and $\gamma=c\sqrt{(\log p)/n}$ in Algorithm \ref{algo2}, the following holds true:
$$\Prob\left\{||\Delta||_{\infty}> \dfrac{c\tilde{c}k_0(\tau k_0+1)}{\phi_0}\cdot\dfrac{\sigma s\log p}{\sqrt{n}} \right\} \leq 2p^{-c'}+2p^{-c''},$$
where  $K=\max_i\sqrt{\widehat{\bSigma}_{i,i}}$ and constants $c'$ and $c''$ are given by
$$c'=\dfrac{\tilde{c}^2}{2K^2}-1,\quad c''=\dfrac{c^2C_{\min}}{24e^2\kappa^4C_{\max}}-2.$$
\end{theorem}

Theorem \ref{th:main} says that  $N(0,\sigma^2\tM\widehat{\bSigma}\tM^\top)$ can be used to approximate the distribution of $\widehat{\beta}^u$ with proper choices of $c$ and $\tilde{c}$ (or equivalently $\gamma$ and $\lambda$). This leads to the following corollary that can be used to  construct asymptotic confidence intervals and p-values for $\beta$ in high-dimensional linear model with linear constraints~\eqref{eq:gmodel0}.
\begin{corollary}\label{th:interval}
Let $\widehat{\sigma}$ be a consistent estimator of $\sigma$.
\begin{enumerate}
\item Define $\delta_i(\alpha,n)=\Phi^{-1}(1-\alpha/2)\widehat{\sigma}n^{-1/2}[\tM\widehat{\bSigma}\tM^\top]_{i,i}^{1/2}$.\\
Then $I_i=[\widehat{\beta}^u_i-\delta_i(\alpha,n), \widehat{\beta}^u_i+\delta_i(\alpha,n)]$ is an asymptotic two-sided level $1-\alpha$ confidence interval for $\beta_i$.
\item For individual hypothesis $H_{0,i}:\beta_{i}=0$ versus $H_{0,i}:\beta_{i}\neq 0$, an asymptotic p-value can be constructed as follows:
$$P_i=2\left[1-\Phi\left(\dfrac{n^{1/2}|\widehat{\beta}^u_i|}{\widehat{\sigma}[\tM\widehat{\bSigma}\tM^\top]_{i,i}^{1/2}}\right)\right].$$
\end{enumerate}
\end{corollary}

The following lemma shows that with Condition 2, the diagonal elements of $\tM\widehat{\bSigma}\tM^\top$ are nonzero with a $\gamma$ that is not too large.
\begin{lemma}\label{th:nonzero}
Let $\tM$ be the matrix obtained by equation~\eqref{eq:tildeM}. Then for $\gamma<(1-(\bP_{\bC})_{i,i})/k_0$ and all $i=1,\dots,p$,
$$[\tM\widehat{\bSigma}\tM^\top]_{i,i}\geq \dfrac{(1-(\bP_{\bC})_{ i,i}-k_0\gamma)^2}{\widehat{\bSigma}_{i,i}}.$$
\end{lemma}

\subsection{Selection of the tuning parameters}\label{tuning}

In real applications, the estimator $\widehat{\beta}^n$, tuning parameter $\lambda$ and estimation of noise level $\widehat{\sigma}$ are obtained through scaled \textsc{lasso} \citep{SunZhang}. Specifically,  the following two steps are iterated until convergence:
\begin{align*}
\widehat{\beta}^n&\leftarrow\argmin_{\bC^\top\beta=0}\left\{||Y-\btZ\beta||_2^2+2n\lambda_0\widehat{\sigma}||\beta||_1\right\},\\
\widehat{\sigma}^2&\leftarrow||Y-\btZ\widehat{\beta}||_2^2/n,
\end{align*}
where $\lambda_0=\sqrt{2}L_n(k/p)$, $L_n(t)=n^{-1/2}\Phi^{-1}(1-t)$, $\Phi^{-1}$ is the quantile function for standard normal and $k$ is the solution of $k=L_1^4(k/p)+2L_1^2(k/p)$. Then $\widehat{\lambda}=\lambda_0\widehat{\sigma}$, 
and $\gamma = a\hat{\lambda}/\hat{\sigma}$ are used in Algorithm 2, where  $a = 1/3$ is used in all simulations and real data analysis in this paper.

\section{Association Between Body Mass Index  and Gut Microbiome}\label{real}
Gut microbiome plays an important role in food  digestion and nutrition absorption. \cite{Wu:Chen:Hoff:Bitt:Chen:Keil:link:2011} reported a cross-sectional study to examine the relationship between micronutrients and gut microbiome composition, where the fecal  samples of 98 healthy volunteers from the University of Pennsylvania were collected, together with demographic data such as body mass index, age and sex.  The DNAs from the fecal samples were analyzed by 454/Roche pyrosequencing of 16S rRNA gene segments of the V1-V2 region. After the pyrosequences were denoised,  a total of about 900,000 16S reads were obtained with an average of 9165 reads per sample and 3068 operational taxonomic units (OTUs) were obtained. These OTUs were combined into 87 genera that appeared in at least one sample. 
Out of these 87 genera, 42 genera have zero counts in more than 90\% of the samples and were removed from our analysis. The remaining 45 relatively common genera belong to four phyla, {\it Actinobacteria, Bacteroidetes, Firmicutes} and {\it Proteobacteria}.  Since dysbiosis of gut microbiome has been shown to be associated with obesity  \citep{Ley:Back:Turn:Lozu:Knig:Gord:obes:2005, Ley:Turn:Klei:Gord:huma:2006, Turn:Ley:Maho:Magr:Mard:Gord:obes:2006}, it is interesting  to identify the bacterial genera 
that are associated with \textsc{bmi} after adjusting for total fat and caloric intakes. In the following analysis, zero count was  replaced by the maximum rounding error of  0.5, commonly used in compositional and microbiome data analysis  \citep{Aitc:stat:2003, Bonneau}.  Since the number of reads is very large, replacing zero with other very small counts does not affect our results. 
These read counts are then converted into compositions of the genera or subcompositions of the genera within phylum.

\subsection{Analysis of the data at the genus-level}
The proposed method was first applied to perform regression analysis with \textsc{bmi} as the response and   the log-transformed  compositions of the 45 genera as the covariates.  In addition,  total fat intake and total caloric intake  were also included as the covariates in the model.  The model was fit with the constraint that the  sum of the coefficients corresponding to the 45 genera  is   zero,  assuming 
$$E(\textsc{BMI})=\sum_{g=1}^{45} \beta_g \log(X_g) + \gamma_1 \textsc{Fat}+\gamma_2 \textsc{Calorie}, $$ 
where  $\sum_{g=1}^{45} \beta_g=0$, and $\log(X_g)$ is the logarithm of the relative abundance of the $g$th genus. The goal of this analysis is to identify the bacteria genera that are associated with  \textsc{bmi}. 

\begin{figure}[htb]
	\centering
	\includegraphics[trim=1cm 7.0cm 2cm 7.0cm, clip=true, width=0.88\linewidth, height=0.35\textheight]{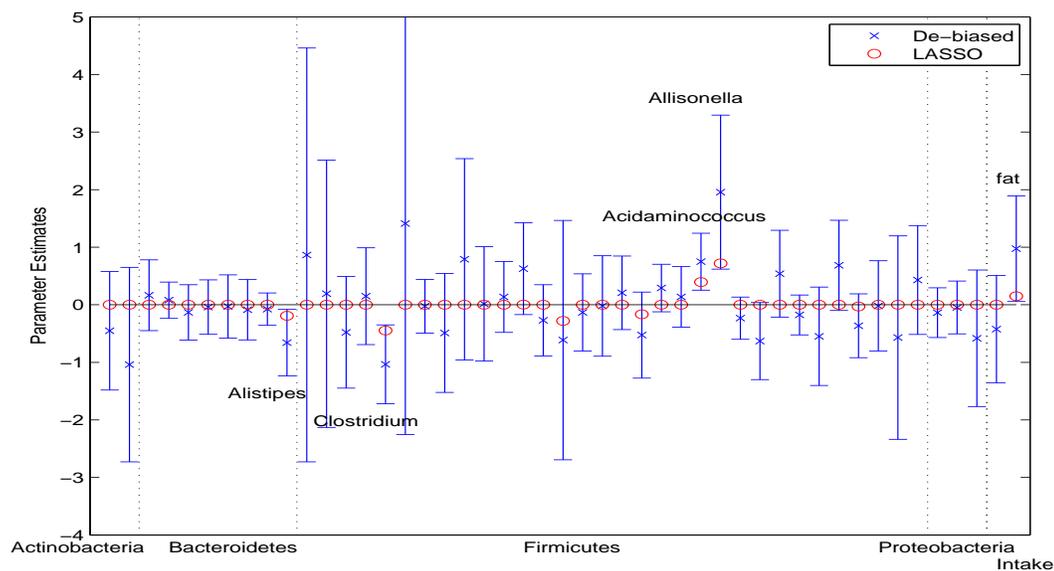}
	\caption{Analysis of gut microbiome data. Lasso estimates, de-biased estimates  and 95\% confidence intervals of the regression coefficients in the model treating the composition of 45 genera as covariates together with total fat and caloric intakes. Dashed vertical lines separate bacterial genus into different phyla. }\label{fig-reg1}
\end{figure}

Figure \ref{fig-reg1} shows the estimated regression coefficients from \textsc{lasso} with one constraint and their de-biased estimates together with the 95\% confidence intervals of the regression coefficients. Four genera were statistically significant with $p$-value of 0.0251 for {\it Alistipes}, 0.0031 for {\it Clostridium}, 0.0031 for {\it Acidaminococcus}, and 0.0042 for {\it Allisonella}, respectively. These  four genera were exactly the same genera  identified using  stability selection by \cite{LinLi}.  They belong to two bacterial phyla,    {\it  Bacteroidetes} and {\it Firmicutes}. The results indicate  that Alistipes in the Bacteroidetes phylum is negatively associated with BMI, which is consistent with previous findings that the gut microbiota in obese mice and humans tend to have a lower proportion of {\it Bacteroidetes} \citep{Ley:Back:Turn:Lozu:Knig:Gord:obes:2005, Ley:Turn:Klei:Gord:huma:2006, Turn:Ley:Maho:Magr:Mard:Gord:obes:2006}. However, for the {\it Firmucutes} phylum,  both the  positively associated ({\it Acidaminococcus} and {\it Allisonella}) and negatively associated ({\it Clostridium}) genera were observed to be associated with \textsc{bmi},  suggesting that obesity may be associated with changes in gut microbiome composition at a lower taxonomic level than previously thought.

\subsection{Subcomposition analysis}
The proposed method was then applied to  subcomposition analysis, where the number of sequencing reads were  converted into compositions of genera within each phylum. This creates four  subcompositions of the genera within four phyla.  This analysis aims to answer the question whether  the composition of genera within a given phylum is associated with \textsc{bmi}, where 
the  log-transformed genera subcompositions are treated  as predictors, together with total fat and caloric intakes as covariates in the following model, 
$$E(\textsc{BMI})=\sum_{g=1}^4 \sum_{s=1}^{m_g} \beta_{gs} \log(X_{gs}) + \gamma_1 \textsc{Fat}+\gamma_2 \textsc{Calorie}, $$ 
where  $\sum_{s=1}^{m_g} \beta_{gs}=0$ for $g=1, \cdots, 4$,  and $\log(X_{gs})$ is the logarithm of the relative abundance of the $s$th genus of the $g$th phylum.   

\begin{figure}[htb]
	\centering
	\includegraphics[trim=1cm 7.0cm 2cm 7.0cm, clip=true, width=0.88\linewidth, height=0.35\textheight]{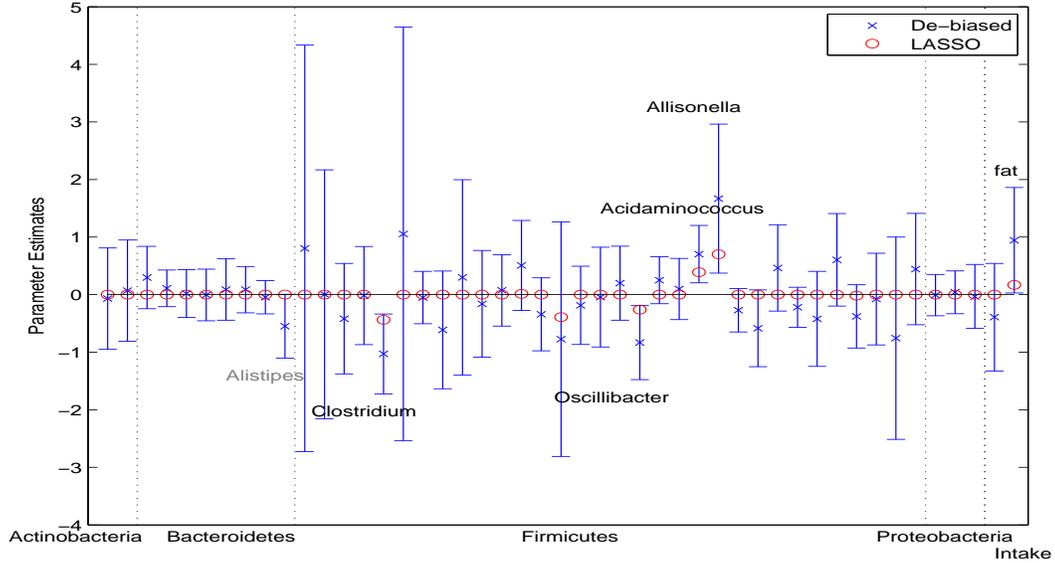}
	\caption{Analysis of gut microbiome data. Lasso estimates, de-biased estimates  and 95\% confidence intervals of the regression coefficients in the model treating the subcompositions of the  genera in each phylum as covariates together with total fat and caloric intakes. Dashed vertical lines separate bacterial genus into different phyla.  }\label{fig-reg2}
\end{figure}

Figure \ref{fig-reg2} shows the \textsc{lasso} estimates, de-biased estimates, and 95\% confidence interval of the coefficients of the 45 genera.
Four genera were statistically significant with $p$-value of 0.0036 for {\it Clostridium}, 0.0056 for  {\it Acidaminococcus}, 0.0116 for {\it  Allisonella}, and 0.0111 for {\it  Oscillibactor}. All  four genera belong to phylum {\it Firmicutes}, indicating that the subcomposition of the bacterial genera within {\it Firmicutes} is associated with \textsc{bmi}.  The genus {\it Alistipes}  has a $p$-value of 0.0523 in this analysis, which is marginally significant.  It is interesting that the bacterial genus {\it Oscillibactor} was identified as one of the two bacterial  genera  that are negatively associated with \textsc{bmi}.  {\it Oscillibacter} was observed to be  increased on the resistent starch and  reduced carbohydrate weight loss diets \citep{Walker2011} in a strictly diet-controlled experiments in obese men, which may explain its negative association with \textsc{bmi}.  Recent study also  identified {\it Oscillibacter}-like organisms as a potentially important gut microbe that mediates high fat diet-induced gut dysfunction \citep{Lam2012}. It is possible that {\it Oscillibacter} directly regulates components involved in the maintenance of gut barrier integrity.

\begin{figure}[hbt]                                                                
 	\centering
\includegraphics[trim=1cm 7.0cm 2cm 7.0cm, clip=true, width=0.88\linewidth, height=0.33\textheight]{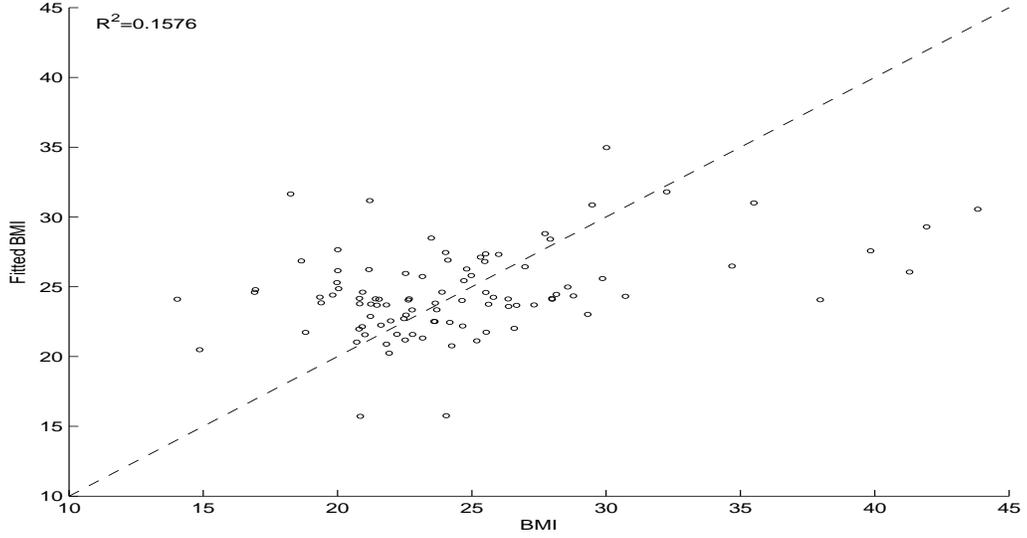}
\caption{Analysis of gut microbiome data. Observed and predicted \textsc{bmi} using LOOCV and variables selected based on  95\% confidence intervals, together with total fat and caloric intakes. }\label{fig-pred2}
\end{figure}

 Figure \ref{fig-pred2} shows the predicted \textsc{bmi} using leave-one-out cross-validation (LOOCV). In each round of LOOCV,  the variables were selected based on the estimated  95\% confidence intervals  and the prediction was performed using refitted  coefficients of the selected bacterial genera, together with calorie and fat intakes.  An $R^2=0.1576$ was obtained between the observed and predicted values. As a comparison, fitting the model with one linear constraint at the genus-level resulted  a $R^2=0.1361$ based on LOOCV, indicating some gain in prediction by the subcompositional analysis.


\section{Simulation Evaluation and Comparisons} \label{simulation}

In order to simulate the compositional covariates, a $n\times p$ matrix $\boldsymbol{W}$ of taxon counts is first generated  with each row of $\boldsymbol{W}$ being generated  from a  log-normal distribution $lnN(\nu,\bSigma)$, where  $\bSigma_{ij} = \zeta^{|i-j|}$ with $\zeta$=0.2 or 0.5 is the covariance matrix to reflect different levels of correlation between the taxa counts.  Parameters  $\nu_j=p/2$ for $j=1,\dots,5$ and $\nu_j=1$ for $j=6,\dots,p$ are set to allow some taxa to be  much more abundant than others, as often observed in real microbiome compositional data.  The compositional covariate matrix $\bZ$ is obtained by normalizing the simulated taxa counts as  $$z_{ij}=\log\left(\frac{w_{ij}}{\sum_{k=1}^p w_{ik}}\right), i=1,\cdots, n, j=1,\dots,p$$. 
Based on these compositional covariates,  the response $Y$ is generated through Model~\eqref{eq:model0} with 
$$\beta=(1, -0.8, 0.4, 0, 0, -0.6, 0, 0, 0, 0, -1.5, 0, 1.2, 0, 0, 0.3, 0, \dots, 0)$$ 
and $\sigma$ = 0.5.
Different   dimension/sample size combinations $(p,n)$=(50,100), (50,200), (50,500), (100,100), (100,200), (100,500) are considered and the simulations are repeated 100 times for each setting.  The tuning parameters are chosen using the method described in Section \ref{tuning}.
The regression coefficient $\beta$ used in the simulation satisfies the following 8 linear constraints 
\begin{equation}\label{eq:constraints}
\begin{split}
&\sum_{j=1}^{10}\beta_j=0, \sum_{j=11}^{16}\beta_j=0, \sum_{j=17}^{20}\beta_j=0, \sum_{j=21}^{23}\beta_j=0, \\ &\sum_{j=24}^{30}\beta_j=0, \sum_{j=31}^{32}\beta_j=0, \sum_{j=33}^{40}\beta_j=0, \sum_{i=41}^p\beta_j=0.
\end{split}
\end{equation}

\subsection{Estimation of confidence intervals}

\begin{figure}[hbt]
	\begin{center}
		\includegraphics[trim=2cm 6.5cm 2cm 7cm, clip=true,width=1\linewidth]{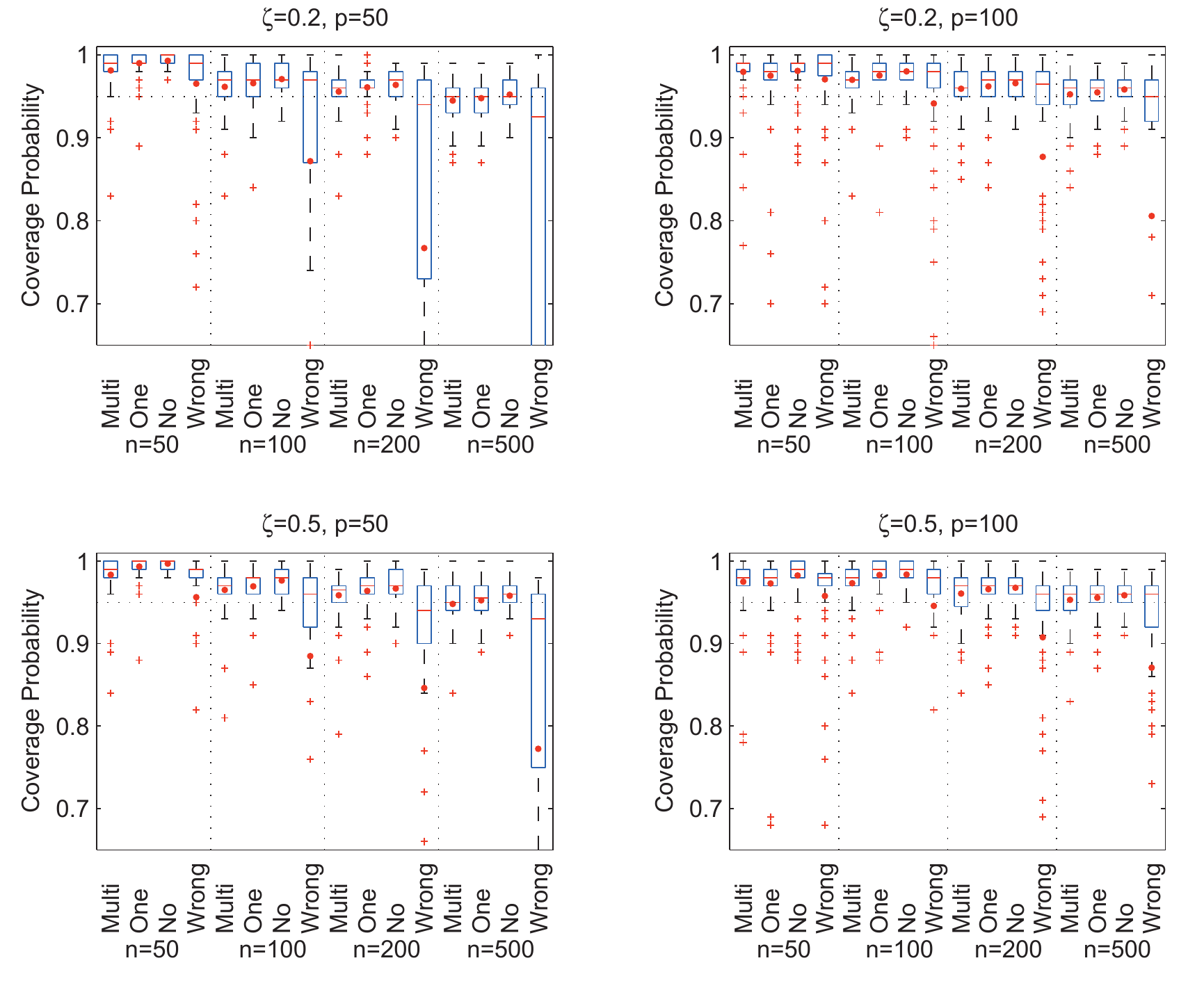}
		\caption{Coverage probabilities of confidence intervals based on 100 replications. For each model, minimum, median (in red line), mean (in red dot) and maximum of the coverage probabilities over compositional covariates  are shown. The boxes labeled with 'c' are obtained  using the constraints, and the boxes labeled with 'u' are obtained  without using  constraints.}\label{fig:cov_prob}
	\end{center}
\end{figure}

\begin{figure}[hbt]
	\begin{center}
		\includegraphics[trim=2cm 6.5cm 2cm 7cm, clip=true,width=1\linewidth]{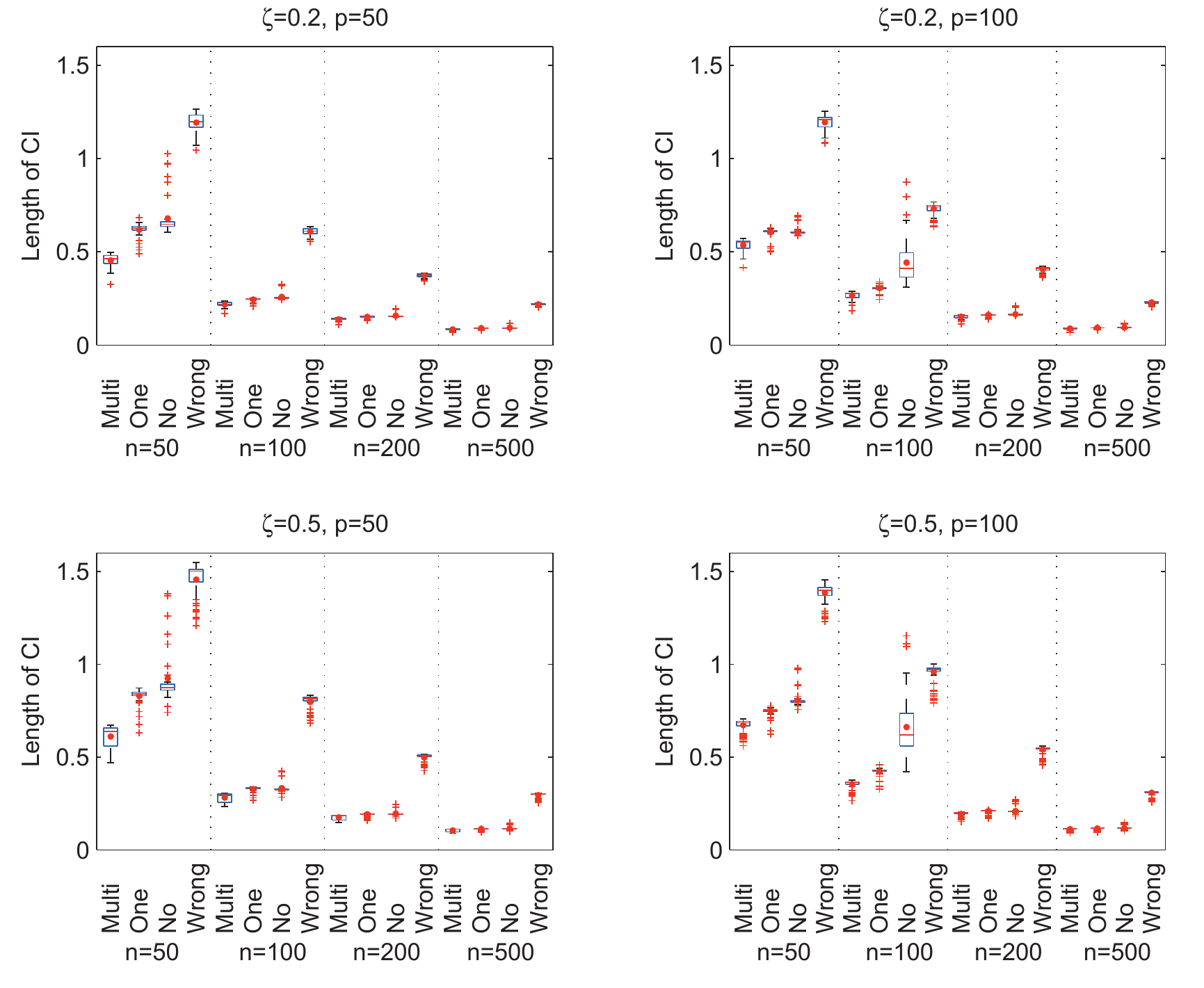}
		\caption{Average lengths of confidence intervals based on 100 replications. For each model, minimum, median (in red line), mean (in red dot) and maximum of the lengths of the intervals overall all compositional covariates are shown.  The boxes labeled with 'c' are obtained  using the constraints, and the boxes labeled with 'u' are obtained  without using  constraints.}\label{fig:CI_length}
	\end{center}
\end{figure}

The model is first fitted under the correct constraints specified in \eqref{eq:constraints} and the corresponding confidence intervals are obtained based on our asymptotic results. Figure \ref{fig:cov_prob} shows the coverage probability for various models and samples sizes, indicating that the coverage probabilities of the confidence intervals are close to the nominal level of 0.95 when the sample size is large. For small sample sizes, the empirical coverage probability  is slightly greater than the nominal level of 0.95, indicating some conservativeness. 
Figure \ref{fig:CI_length} shows the lengths of confidence intervals. As expected,  larger sample sizes result in shorter lengths and larger correlations among the variables lead to increased length of the confidence intervals. 

As comparisons, the model is also fitted under no constraint, one single constraint, $\sum_{j=1}^p\beta_j=0$, and misspecified constraints, 
\begin{equation*}
\sum_{j=1}^{5}\beta_j=0, \sum_{j=6}^{12}\beta_j=0, \sum_{j=13}^{23}\beta_j=0, \sum_{j=24}^{30}\beta_j=0,\sum_{j=31}^{p}\beta_j=0.
\end{equation*}	
The coverage probabilities and the lengths of the confidence intervals are given in Figure \ref{fig:cov_prob} and Figure \ref{fig:CI_length}, respectively.  While the coverage probabilities are relatively less sensitive to such misspecification, 
the intervals estimated under the correct linear constraints are much shorter than those obtained with one or none of the linear constraints, especially when sample size is small. Using the wrong constraints  results in  much longer intervals with less accurate coverage.

\subsection{Variable selection based on the confidence intervals}
The confidence intervals of the regression coefficients can also be applied to choose the variables of interest.  For example, a variable can be selected if the  nominal $95\%$ confidence interval of the corresponding regression coefficient  includes zero. Table~\ref{tab:TPFP} shows the true positive rate and false positive rate of the variables identified based on  $95\%$ confidence intervals under multiple constraints, one single constraint and no constraint.  When the sample size is small, imposing the correct linear constraints can lead to more true discoveries  while the false positive rates are still controlled under $5\%$. In contrast, the models with only one or no constraint 
lead to much lower true positive rates and the standard \textsc{lasso}  without any constraint gives the worst
variable selection results.

\begin{table}[htb]
	\begin{center}
		\caption{True/False positive rates of the significant variables selected based on $95\%$  confidence intervals with multiple, one and no linear constraints. Variable correlations $\zeta$, numbers of variables $p$ and sample sizes ($n$) are considered. }\label{tab:TPFP}
		\begin{tabular}{ccccccccc}
			\hline
			\multicolumn{3}{c}{Configuration}& \multicolumn{3}{c}{True Positive Rate} & \multicolumn{3}{c}{False Positive Rate} \\
			\cmidrule(lr){1-3} \cmidrule(lr){4-6}  \cmidrule(lr){7-9}
			\multicolumn{3}{c}{}& \multicolumn{3}{c}{Constraints} & \multicolumn{3}{c}{Constraints} \\
			$\zeta$ & $p$ & $n$ & Multi & One & No & Multi & One & No\\
			\hline
			\multirow{4}{*}{0.2}&\multirow{4}{*}{50}
			& 50 & 0.9329&0.8514&0.7586&0.0121&0.0056&0.0051\\
			&& 100 & 1.0000&1.0000&0.9957&0.0330&0.0286&0.0267\\
			&& 200 & 1.0000&1.0000&1.0000&0.0386&0.0333&0.0328\\
			&& 500 & 1.0000&1.0000&1.0000&0.0498&0.0477&0.0470\\
			\hline
			\multirow{4}{*}{0.2}&\multirow{4}{*}{100}
			& 50 & 0.8571&0.8071&0.7700&0.0131&0.0166&0.0139\\
			&& 100 & 1.0000&0.9857&0.9400&0.0265&0.0218&0.0173\\
			&& 200 & 1.0000&1.0000&1.0000&0.0374&0.0353&0.0333\\
			&& 500 & 1.0000&1.0000&1.0000&0.0441&0.0428&0.0406\\
			\hline
			\multirow{4}{*}{0.5}&\multirow{4}{*}{50}
			& 50 & 0.8500&0.7486&0.6543&0.0095&0.0030&0.0019\\
			&& 100 & 0.9971&0.9900&0.9871&0.0281&0.0240&0.0223\\
			&& 200 & 1.0000&1.0000&1.0000&0.0351&0.0309&0.0305\\
			&& 500 & 1.0000&1.0000&1.0000&0.0474&0.0437&0.0412\\
			\hline
			\multirow{4}{*}{0.5}&\multirow{4}{*}{100}
			& 50 & 0.7643&0.7157&0.6443&0.0168&0.0173&0.0118\\
			&& 100 & 0.9814&0.9300&0.8500&0.0227&0.0137&0.0145\\
			&& 200 & 1.0000&1.0000&1.0000&0.0359&0.0320&0.0319\\
			&& 500 & 1.0000&1.0000&1.0000&0.0444&0.0417&0.0409\\
			\hline
		\end{tabular}
		
	\end{center}
\end{table}

\subsection{Prediction evaluation}
Prediction performances are also evaluated and compared for models with or without linear constraints.  The prediction error $||Y-\bZ\hat{\beta}||_2^2/n$ is  computed from an independent test sample of size $n$. 
	Table~\ref{tab:pred_error} shows the prediction errors of the \textsc{lasso} estimator, refitted estimator with variables selected by  \textsc{lasso}, and refitted estimator with variables selected by the 95\% confidence intervals. For  each of these three estimators,  model fitting and coefficient refitting and prediction are performed with multiple, one and no linear constraints. Overall, fitting the models with correct multiple  constraints substantially decreases the prediction error. The \textsc{lasso} estimator has the worst prediction performance, while the two refitted estimators have comparable prediction errors.

\begin{table}[htb]
	\begin{center}
		\caption{Testing set prediction error of the \textsc{lasso} estimator, refitted estimator with variables selected by  by \textsc{lasso}, and refitted estimator with variables selected based on $95\%$  confidence intervals.  For each estimator, model was fit using  multiple, one and no linear constraints. Variable correlations $\zeta$, numbers of variables $p$ and sample sizes ($n$) are considered. }\label{tab:pred_error}
		\setlength{\tabcolsep}{5pt}
		\begin{tabular}{cccccccccccc}
			\hline
			\multicolumn{3}{c}{}& \multicolumn{3}{c}{} & \multicolumn{3}{c}{Refitted with} & \multicolumn{3}{c}{Refitted with}\\
			\multicolumn{3}{c}{Configuration}& \multicolumn{3}{c}{\textsc{lasso} Estimator} & \multicolumn{3}{c}{Selection by \textsc{lasso}} & \multicolumn{3}{c}{Selection by 95\% CI}\\
			\cmidrule(lr){1-3} \cmidrule(lr){4-6}  \cmidrule(lr){7-9} \cmidrule(lr){10-12}
			\multicolumn{3}{c}{} & \multicolumn{3}{c}{Constraints} & \multicolumn{3}{c}{Constraints} & \multicolumn{3}{c}{Constraints} \\
			$\zeta$ & $p$ & $n$ & Multi & One & No & Multi & One & No & Multi & One & No\\
			\hline
			\multirow{4}{*}{0.2}&\multirow{4}{*}{50}
			& 50 & 0.687&0.926&0.983&0.360&0.502&1.336&0.370&0.487&1.375\\
			&& 100 & 0.360&0.391&0.412&0.300&0.309&1.153&0.284&0.296&1.155\\
			&& 200 & 0.293&0.302&0.307&0.271&0.273&1.039&0.264&0.269&1.054\\
			&& 500 & 0.265&0.269&0.270&0.259&0.261&1.025&0.255&0.258&1.034\\
			\hline
			\multirow{4}{*}{0.2}&\multirow{4}{*}{100}
			& 50 & 1.027&1.429&1.438&0.484&0.776&1.531&0.496&0.602&1.483\\
			&& 100 & 0.408&0.467&0.491&0.305&0.315&1.164&0.286&0.322&1.300\\
			&& 200 & 0.303&0.318&0.322&0.273&0.276&1.066&0.268&0.277&1.076\\
			&& 500 & 0.269&0.274&0.274&0.263&0.264&1.041&0.260&0.264&1.049\\
			\hline
			\multirow{4}{*}{0.5}&\multirow{4}{*}{50}
			& 50 & 0.806&1.095&1.210&0.520&0.687&1.179&0.441&0.557&1.278\\
			&& 100 & 0.400&0.476&0.454&0.300&0.319&0.959&0.283&0.301&0.963\\
			&& 200 & 0.305&0.325&0.320&0.270&0.272&0.861&0.263&0.267&0.877\\
			&& 500 & 0.269&0.276&0.274&0.258&0.260&0.847&0.255&0.257&0.862\\
			\hline
			\multirow{4}{*}{0.5}&\multirow{4}{*}{100}
			& 50 & 1.069&1.494&1.731&0.668&0.993&1.416&0.606&0.690&1.361\\
			&& 100 & 0.476&0.604&0.560&0.322&0.366&0.963&0.293&0.342&1.134\\
			&& 200 & 0.323&0.358&0.342&0.271&0.273&0.884&0.265&0.270&0.896\\
			&& 500 & 0.274&0.284&0.279&0.262&0.262&0.863&0.258&0.261&0.876\\
			\hline
		\end{tabular}
					\normalsize
		
	\end{center}
\end{table}

\subsection{Simulation based on real microbiome compositional data}

\begin{figure}
	\centering
	\includegraphics[trim=0cm 10cm 0cm 10cm, clip=true,width=1\linewidth]{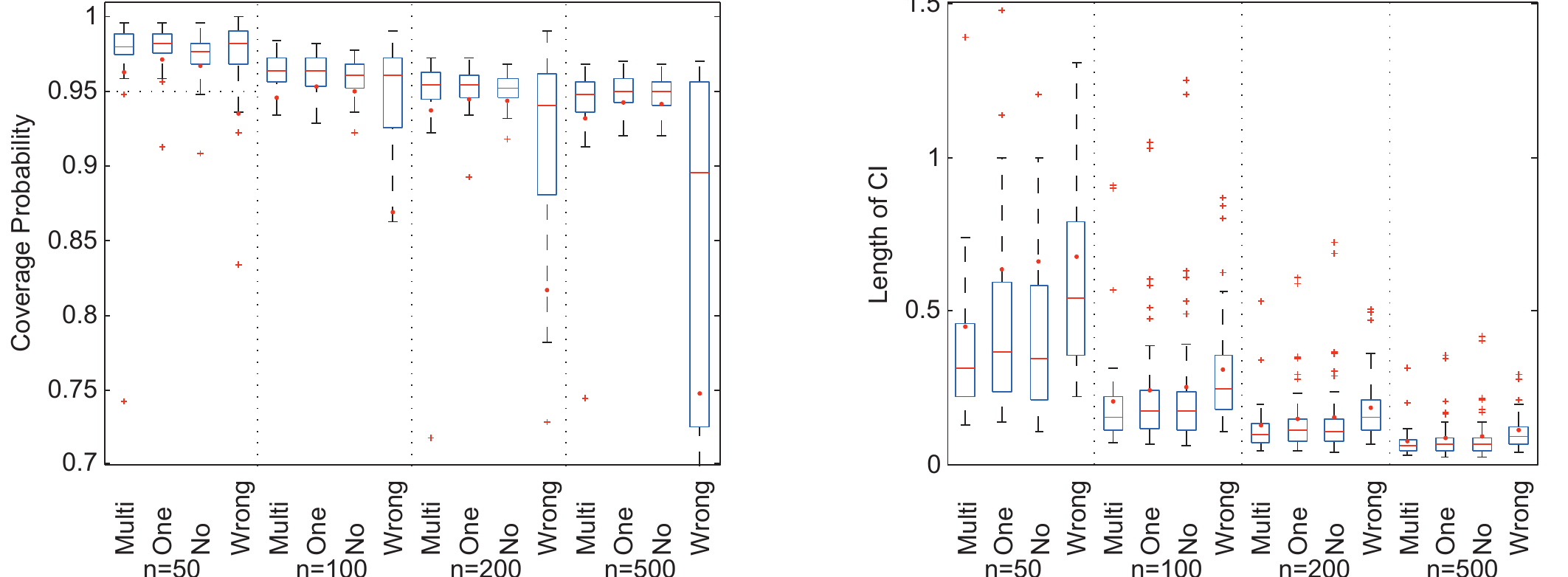}
	\caption{Coverage probabilities and length of confidence intervals based on 500 replications. Data are simulated by resampling the  gut microbiome composition data in Section \ref{real}.}
	\label{fig:simu_BMI_v2}
\end{figure}

Another set of simulations are conducted where  the gut microbiome composition data analyzed in Section \ref{real}  are used to generate  the covariates with  $p=45$ through resampling. The many zeros in the compositional data matrix are replaced with pseudo-count of 0.05 and are renormalized to have unit sum. For each simulation, we resample with replacement from the rows of compositional data matrix to achieve the required sample size. The coefficients $\beta$ and noise level $\sigma$ are the same as in prevision section.  The sample size is   chosen to be $n=50, 100, 200$ and $500$. Each setting is repeated 500 times.  The coverage probability and length of confidence intervals are shown in Figure~\ref{fig:simu_BMI_v2} for model with multiple, one and no constraints on the coefficients.   Similar conclusions are observed. The coverage probabilities are relatively less sensitive to  misspecification of linear constraints, however,  
the intervals estimated under the correct linear constraints are shorter than those obtained with one or none of the linear constraints, especially when sample size is small. Using the wrong constraints  results in  much longer intervals with a less accurate coverage.

\section{Discussion}
This paper  has considered the problem of regression analysis for microbiome compositional data obtained through 16S sequencing or metagenomic sequencing.   
The models and methods in this paper can be applied to identify the microbial subcompositions that are associated with a continuous response.  The idea of imposing the constraints on regression coefficients was motivated by using the log-ratios as covariates. However, the method proposed does not use the log-ratios as covariates, it treats the logarithm of  the relative abundances as covariates and allows the response to depend on the relative abundances of certain bacteria
instead of the ratios. Imposing linear constraints on coefficients enhances the interpretability and also guarantees the subcompositional coherence.  Our method allows selecting taxa in different higher rank taxa.  By applying our subcompositional analysis,  {\it Oscillibacter} genus was found to be  associated with \textsc{bmi}, even after total fat and caloric intakes were adjusted, indicating that gut microbiome may serve as independent predictor for complex phenotypes such as $\textsc{bmi}$.  Our simulation studies have demonstrated a clear gain in prediction performance when true linear constraints are imposed. However, the small sample size of our data did not allow us to extensively evaluate gain in $\textsc{bmi}$ prediction by incorporating the gut microbiome data.

  An estimation procedure through   regularization  under linear constraints has been developed.   In order to obtain the confidence interval of the regression coefficients,   de-biased estimates of the regression coefficients  are obtained, which are shown to be  approximately   normally distributed.    The $p$ optimization problems in the de-biased algorithm can be solved efficiently using convex programs. For one simulated data set in Section \ref{simulation}, Algorithm 2 took about 36 seconds for $p=100$  and 300 seconds for $p=200$ on a PC with a core of Intel i7-3770 CPU\@3.40GHz. For large $p$,  convex optimization problems can be carried out in parallel. In typical microbiome studies,   $p$ is less than 1,000.

   The general results presented in this paper can also be used for statistical inference for the log-contrast model considered in \cite{LinLi}.   This type of de-biased estimates were also proposed in 
  \cite{ZhangZhang} and \cite{vanderGeer2014}. \cite{LeeTaylor2014} proposed an exact inference procedure for \textsc{lasso} by  characterizing the distribution of a post-selection estimator conditioned on the selection event. It is interesting to extend their approach to the high-dimensional regression problems with constraints.  
  \cite{Efron2014JASA}  developed a bootstrap smoothing  procedure for computing the standard errors and confidence intervals for predictions, which is different from what was considered in this paper.  Efron's procedure can be applied directly to make  inferences on predictions using the methods developed here.  
  
 Several extensions are worth considering.  Model \eqref{eq:gmodel0} can be extended to include the interaction terms  of the form $\lambda_{lk}(\log x_{il}-\log x_{ik})^2$, where $x_{il}$ and $x_{ik}$ are the proportion of the $l$th and the $k$th component of subject $i$, $\lambda_{lk}$ is the coefficient that corresponds to the interaction between these two components \citep{Aitc:Baco:log:1984}.  Similar variable selection and inference procedure can be developed. It is also interesting to develop  methods for  generalized linear models with high-dimensional compositional data as covariates.


\section*{Acknowledgments}
We are grateful to the editor, the associate editor, and three anonymous referees for their helpful comments.
We thank Drs. Gary Wu, James Lews and Rick Bushman for  sharing the data and Dr. Wei Lin for helpful discussions. 


\bibliographystyle{apa}
\bibliography{comp_sel}

\section*{Appendix}

We collect the proofs of the main results in this Appendix.

\begin{definition}
	For any matrix $M$, define the restricted orthogonal constant (ROC) of order $k_1$ and $k_2$ as below
	\begin{align*}
	\theta_{k_1,k_2}(M)=\sup\Big\{\dfrac{|\langle M\alpha_1,M\alpha_2\rangle|}{||\alpha_1||_2||\alpha_2||_2}:&\alpha_1 \mbox{ is } k_1\mbox{-sparse vector}, \alpha_2 \mbox{ is } k_2\mbox{-sparse vector},\\
	& \alpha_1 \mbox{ and } \alpha_2 \mbox{ have non-overlapping support}\Big\}
	\end{align*}
\end{definition}

Before proving Theorem \ref{th:main}, we need to present Lemma \ref{th:proj} and Theorem \ref{th:main0}.
\begin{lemma}\label{th:proj}
	Suppose $||\biden - P_C||_{\infty}\leq k_0$, then for any matrix $A$, we have
	$$|(\biden-P_C)A|_\infty\leq k_0|A|_\infty.$$
\end{lemma}
{\bf\noindent Proof of Lemma \ref{th:proj}.}
	By definition of $||\cdot||_\infty$ for matrices, for any vector $a\in\mathbb{R}^p$,
	$$||(\biden-P_C) a||_{\infty}\leq k_0||a||_{\infty}.$$
$\square$

\begin{theorem}\label{th:main0}
	Let $\widehat{\beta}^n$ be the estimator obtained by solving optimization problem~\eqref{eq:l1opt} for Model~\eqref{eq:gmodel},
	where $\beta$ is $s$-sparse.
	If $(3\tau-1)\delta^{-}_{2s}(\btZ/\sqrt{n})-(\tau+1)\delta^{+}_{2s}(\btZ/\sqrt{n})\geq 4\tau\phi_0$ for some constant $\phi_0>0$, and $||\btZ^\top\epsilon||_{\infty}\leq n\lambda/\tau$,
	then,
	$$||\widehat{\beta}^n-\beta||_1\leq s\lambda(k_0+1/\tau)/\phi_0.$$
\end{theorem}

{\noindent\bf Proof.}
	By the definition of $\widehat{\beta}^n$, we have
	$$\dfrac{1}{2n}||y-\btZ\widehat{\beta}^n||_2^2+\lambda||\widehat{\beta}^n||_1\leq \dfrac{1}{2n}||y-\btZ\beta||_2^2+\lambda||\beta||_1.$$
	Denote $h=\widehat{\beta}^n-\beta$, and $S_h$ be the set of index of the $s$ largest absolute values of $h$.\\
	Then by $Y=\btZ\beta+\epsilon$, we have
	\begin{equation}\label{eq:ineq1}
	\dfrac{1}{2n}(||\epsilon-\btZ h||_2^2-||\epsilon||_2^2)\leq \lambda(||\beta||_1-||\widehat{\beta}^n||_1).
	\end{equation}
	Notice that
	\begin{align*}
	||\beta||_1-||\widehat{\beta}^n||_1 &= ||\beta_{supp(\beta)}||_1-||\widehat{\beta}^n_{supp(\beta)}||_1-||\widehat{\beta}^n_{supp(\beta)^c}||_1\\
	&\leq ||\beta_{supp(\beta)}-\widehat{\beta}^n_{supp(\beta)}||_1-||h_{supp(\beta)^c}||_1\\
	&\leq ||h_{supp(\beta)}||_1-||h_{supp(\beta)^c}||_1\\
	&\leq ||h_{S_h}||_1-||h_{S_h^c}||_1.
	\end{align*}
	Also,
	\begin{align*}
	\dfrac{1}{2n}(||\epsilon-\btZ h||_2^2-||\epsilon||_2^2)&=-\dfrac{1}{2n}(\btZ h)^\top (2\epsilon -\btZ h)\geq -\dfrac{1}{n}h^\top \btZ^\top \epsilon
	\geq -\dfrac{1}{n} ||\btZ^\top \epsilon||_{\infty}||h||_1\\
	&= -\dfrac{1}{n} ||\btZ^\top \epsilon||_{\infty}(||h_{S_h}||_1+||h_{S_h^c}||_1).
	\end{align*}
	Then, by $||\btZ^\top\epsilon||_{\infty}\leq n\lambda/\tau$ and~\eqref{eq:ineq1}, we have
	$$ -(||h_{S_h}||_1+||h_{S_h^c}||_1)\leq \tau(||h_{S_h}||_1-||h_{S_h^c}||_1).$$
	Therefore,
	\begin{equation}\label{eq:ineq2}
	||h_{S_h^c}||_1 \leq \dfrac{\tau+1}{\tau-1}||h_{S_h}||_1.
	\end{equation}
	By the KKT condition of optimization problem~\eqref{eq:l1opt}, we have $$||\btZ^\top(y-\btZ\widehat{\beta}^n)+C\boldsymbol{\mu}||_\infty\leq n\lambda$$
	for some $\boldsymbol{\mu}\in\mathbb{R}^r$.
	Then by Lemma \ref{th:proj},
	\begin{align*}
	||\btZ^\top(y-\btZ\widehat{\beta}^n)||_\infty &= ||(\biden-P_C)(\btZ^\top(y-\btZ\widehat{\beta}^n)+C\boldsymbol{\mu})||_\infty\\
	&\leq k_0||\btZ^\top(y-\btZ\widehat{\beta}^n)+C\boldsymbol{\mu}||_\infty
	\leq k_0n\lambda.
	\end{align*}
	Then
	$$||\btZ^\top\btZ h||_\infty\leq||\btZ^\top(y-\btZ\widehat{\beta}^n)||_\infty + ||\btZ^\top(y-\btZ\beta)||_\infty\leq k_0n\lambda+||\btZ^\top\epsilon||_\infty.$$
	Using Lemma 5.1 in \cite{Cai:Zhang:RIP:2013}, we can get
	\begin{align*}
	|\langle\btZ h_{S_h}, \btZ h_{S_h^c} \rangle|&\leq \theta_{s,s}(\btZ)||h_{S_h}||_2\cdot\max(||h_{S_h^c}||_{\infty},||h_{S_h^c}||_1/s)\sqrt{s}\\
	&\leq\sqrt{s}\theta_{s,s}(\btZ)||h_{S_h}||_2\cdot \dfrac{\tau+1}{\tau-1}||h_{S_h}||_1/s\\
	&\leq\dfrac{\tau+1}{\tau-1}\theta_{s,s}(\btZ)||h_{S_h}||_2^2.
	\end{align*}
	Then,
	\begin{align*}
	(k_0 n\lambda+||\btZ^\top\epsilon||_\infty)||h_{S_h}||_1&\geq ||\btZ^\top\btZ h||_\infty||h_{S_h}||_1 \geq \langle \btZ^\top\btZ h, h_{S_h} \rangle\\
	&=\langle \btZ h_{S_h}, \btZ h_{S_h} \rangle + \langle \btZ h_{S_h}, \btZ h_{S_h^c} \rangle\\
	&\geq ||\btZ h_{S_h}||_2^2 - \dfrac{\tau+1}{\tau-1}\theta_{s,s}(\btZ)||h_{S_h}||_2^2\\
	&\geq(\delta_{2s}^-(\btZ)-\dfrac{\tau+1}{\tau-1}\theta_{s,s}(\btZ))||h_{S_h}||_2^2\\
	&\geq\left(\dfrac{3\tau-1}{2(\tau-1)}\delta_{2s}^-(\btZ)-\dfrac{\tau+1}{2(\tau-1)}\delta_{2s}^+(\btZ)\right)||h_{S_h}||_1^2/s.
	\end{align*}
	The last inequality comes from $\theta_{k_1,k_2}(A)\leq \dfrac{1}{2}(\delta^{+}_{k_1+k_2}(A)-\delta^{-}_{k_1+k_2}(A))$ for any matrix $A$ from Lemma 1 of \cite{Kang:Zhang:instru:2014}.
	The inequality above gives us
	$$||h_{S_h}||_1\leq s\dfrac{k_0n\lambda+||\btZ^\top\epsilon||_\infty}{\dfrac{n}{2(\tau-1)}\left((3\tau-1)\delta_{2s}^-(\btZ/\sqrt{n})-(\tau+1)\delta_{2s}^+(\btZ/\sqrt{n})\right)} = s\dfrac{k_0n\lambda+||\btZ^\top\epsilon||_\infty}{2n\tau\phi_0/(\tau-1)}.$$
	Therefore, by $||\btZ^\top\epsilon||_{\infty}\leq n\lambda/\tau$ and~\eqref{eq:ineq2}, we have
	$$||\widehat{\beta}^n-\beta||_1=||h_{S_h}||_1+||h_{S_h^c}||_1\leq \dfrac{2\tau}{\tau-1}||h_{S_h}||_1\leq s\lambda(k_0+1/\tau)/\phi_0.$$
$\square$

{\noindent\bf Proof of Theorem~\ref{th:main}.}
	\begin{align*}
	\widehat{\beta}^u-\beta&=\widehat{\beta}^n-\beta+\dfrac{1}{n}\tM\btZ^\top\epsilon + \dfrac{1}{n}\tM\btZ^\top\btZ(\beta-\widehat{\beta}^n)\\
	&=\dfrac{1}{n}\tM\btZ^\top\epsilon + (\tM\widehat{\Sigma}-\biden)(\beta-\widehat{\beta}^n)\\
	&=\dfrac{1}{n}\tM\btZ^\top\epsilon + (\tM\widehat{\Sigma}-\biden + P_C)(\beta-\widehat{\beta}^n),
	\quad \mbox{ since $C^\top\beta=C^\top\widehat{\beta}^n=0$.}
	\end{align*}
	Thus, $\sqrt{n}(\widehat{\beta}^u-\beta)=B+\Delta$ where $B=\dfrac{1}{\sqrt{n}}\tM\btZ^\top\epsilon$.\\
	Notice that $(\biden-P_C)\widehat{\Sigma}(\biden-P_C)=\widehat{\Sigma}$, and $B=\frac{1}{\sqrt{n}}\tM\btZ^\top\epsilon=\frac{1}{\sqrt{n}}\tM (\biden-P_C)\bZ^\top\epsilon$.
	Thus,
	$$B|\bZ\sim N\left(0, \sigma^2\tM (\biden-P_C)\widehat{\Sigma}(\biden-P_C)\tM^\top \right)=N(0, \sigma^2\tM\widehat{\Sigma}\tM^\top).$$
	\begin{align*}
	||\Delta||_{\infty}&\leq \sqrt{n}\left|\tM\widehat{\Sigma}-(\biden-P_C)\right|_{\infty}||\beta-\widehat{\beta}^n||_1\\
	&= \sqrt{n}\left|(\biden-P_C)\left(M\widehat{\Sigma}-(\biden-P_C)\right)\right|_{\infty}||\beta-\widehat{\beta}^n||_1\\
	&\leq k_0\sqrt{n}\left|M\widehat{\Sigma}-(\biden-P_C)\right|_{\infty}||\beta-\widehat{\beta}^n||_1.
	\end{align*}
	The last inequality is by Lemma \ref{th:proj}.\\
	By Lemma \ref{th:feasible}, when choosing $\gamma=c\sqrt{(\log p)/n}$, $\Omega$ is a feasible solution of the optimization problem~\eqref{eq:opt} with probability at least $1-2p^{-c''}$. Therefore, $\left|M\widehat{\Sigma}-(\biden-P_C)\right|_{\infty}\leq \gamma=c\sqrt{(\log p)/n}$ with probability at least $1-2p^{c''}$.\\
	By Theorem \ref{th:main0}, take $\lambda = \tau\tilde{c}\sigma\sqrt{(\log p)/n}$,
	\begin{align*}
	\Prob(||\widehat{\beta}^n-\beta||_1\leq (k_0+1/\tau)\lambda s/\phi_0)
	&\geq 1- \Prob (||\btZ^\top \epsilon||_\infty > n\lambda/\tau)\\
	& \geq 1 - \sum_{i=1}^p \Prob(|(\btZ^\top \varepsilon)_i| > n\lambda/\tau)\\
	&\geq 1 -  2p \exp\left\{-\dfrac{1}{2}\dfrac{(n\lambda/\tau)^2}{n(\sigma K)^2}\right\}\\
	&=1 - 2p^{1-\tilde{c}^2/(2K^2)}=1-2p^{-c'}.
	\end{align*}
	Altogether, we have
	\begin{eqnarray*}
		&&\Prob\left\{||\Delta||_{\infty}>\dfrac{c\tilde{c}k_0(k_0\tau+1)}{\phi_0}\dfrac{s\sigma\log p}{\sqrt{n}}  \right\}\\
		&\leq&
		\Prob\left\{||\widehat{\beta}^n-\beta||_1\leq s\lambda(k_0+1/\tau)/\phi_0 = \tilde{c}(k_0\tau+1)s\sigma\sqrt{(\log p)/n}/\phi_0\right\} \\
		&+& \Prob\left\{\left|M\widehat{\Sigma}-(\biden-P_C)\right|_{\infty}\leq \gamma=c\sqrt{(\log p)/n}\right\}\\
		&\leq& 2p^{-c'}+2p^{-c''}.
	\end{eqnarray*}
$\square$

{\noindent\bf Proof of Lemma~\ref{th:feasible}.}
	Note that $\Sigma^{1/2}\Omega^{1/2}\widetilde{Z}_l=(\biden-P_C)\widetilde{Z}_l=\widetilde{Z}_l$. Therefore,
	\begin{align*}
	\Omega\widehat{\Sigma}-(\biden-P_C)&=\dfrac{1}{n}\sum_{l=1}^n\left\{\Omega\widetilde{Z}_l\widetilde{Z}_l^\top-(\biden-P_C)
	\right\}\\
	&=\dfrac{1}{n}\sum_{l=1}^n\left\{\Omega^{1/2}\Omega^{1/2}\widetilde{Z}_l\widetilde{Z}_l^\top
	\Omega^{1/2}\Sigma^{1/2}
	-(\biden-P_C)
	\right\}.
	\end{align*}
	Define $v_l^{(ij)}=\Omega^{1/2}_{i,\cdot}\Omega^{1/2}\widetilde{Z}_l\widetilde{Z}_l^\top
	\Omega^{1/2}\Sigma^{1/2}_{\cdot,j} - (\biden-P_{C})_{i,j}$.
	Since $\Exp \Omega\widetilde{Z}_l\widetilde{Z}_l^\top=
	\Omega\Sigma=(\biden-P_C)$, we have
	$\Exp v_l^{(ij)} = 0$. Then, by the proof of Lemma 6.2 in \cite{Javanmard:2014},
	\begin{align*}
	||v_l^{(ij)}||_{\psi_1}&\leq 2||\Omega^{1/2}_{i,\cdot}\Omega^{1/2}\widetilde{Z}_l\widetilde{Z}_l^\top
	\Omega^{1/2}\Sigma^{1/2}_{\cdot,j}||_{\psi_1}\\
	&\leq 2||\Omega^{1/2}_{i,\cdot}\Omega^{1/2}\widetilde{Z}_l||_{\psi_2}||\Sigma^{1/2}_{j,\cdot}\Omega^{1/2}\widetilde{Z}_l||_{\psi_2}\\
	&\leq 2||\Omega^{1/2}_{i,\cdot}||_2||\Sigma^{1/2}_{j,\cdot}||_2||\Omega^{1/2}\widetilde{Z}_l||_{\psi_2}||\Omega^{1/2}\widetilde{Z}_l||_{\psi_2}\\
	&\leq 2\sqrt{\sigma_{\max}(\Sigma)\sigma_{\max}(\Omega)}\kappa^2\\
	&\leq  2\sqrt{C_{\max}/C_{\min}}\kappa^2\equiv \kappa',
	\end{align*}
	where $||X||_{\psi_1}$ is the sub-exponential norm of a random variable $X$ and is defined as
	$$||X||_{\psi_1} = \sup_{p\geq 1}p^{-1}(\Exp|X|^p)^{1/p}.$$
	Applying Bernstein-type inequality for centered sub-exponential random variables \citep{Buhl:Geer:stat:2011}, we get
	$$\Prob\left\{\dfrac{1}{n}\left|\sum_{l=1}^n v_l^{(ij)}\right|\geq \gamma \right\} \leq 2\exp\left[-\dfrac{n}{6}\min\left((\dfrac{\gamma}{e\kappa'})^2,\dfrac{\gamma}{e\kappa'}\right)\right].$$
	Take $\gamma = c\sqrt{(\log p)/n}$ with $c\leq e\kappa'\sqrt{n/\log p}$, we have
	$$\Prob\left\{\dfrac{1}{n}\left|\sum_{l=1}^n v_l^{(ij)}\right|\geq c\sqrt{\dfrac{\log p}{n}} \right\} \leq 2p^{-c^2/(6e^2\kappa'^2)}=2p^{-(c^2C_{\min})/(24e^2\kappa^4C_{\max})}.$$
	Therefore, by union bounding over all pairs of $i$ and $j$,
	$$\Prob\left\{\left|\Omega\widehat{\Sigma}-(\biden-P_C)\right|_{\infty}\geq c\sqrt{\dfrac{\log p}{n}} \right\} \leq 2p^{-(c^2C_{\min})/(24e^2\kappa^4C_{\max})+2}.$$
$\square$

{\noindent\bf Proof of Lemma \ref{th:nonzero}.}
	Suppose $\tM=(\widetilde{m}_1, \dots, \widetilde{m}_p)^\top$. Since
	$$\left|\tM\widehat{\Sigma}-(\biden-P_C)\right|_{\infty}\leq k_0\left|M\widehat{\Sigma}-(\biden-P_C)\right|_{\max}\leq k_0\gamma,$$
	we have
	$1-P_{C i,i}-e_i^\top\widehat{\Sigma}\widetilde{m}_i\leq k_0\gamma$.
	Therefore, for all $L\geq 0$,
	\begin{align*}
	\widetilde{m}_i^\top\widehat{\Sigma}\widetilde{m}_i&\geq \widetilde{m}_i^\top\widehat{\Sigma}\widetilde{m}_i + L(1-P_{C i,i}-k_0\gamma)-Le_i^\top\widehat{\Sigma}\widetilde{m}_i\\
	&\geq \min_{m}\left\{m^\top\widehat{\Sigma}m + L(1-P_{C i,i}-k_0\gamma)-Le_i^\top\widehat{\Sigma}m\right\}\\
	&= L(1-P_{C i,i}-k_0\gamma)-\dfrac{L^2}{4}\widehat{\Sigma}_{i,i} \quad \mbox{(The minimizer $m=Le_i/2$)}\\
	&\geq \min_{L\geq0}\left\{L(1-P_{C i,i}-k_0\gamma)-\dfrac{L^2}{4}\widehat{\Sigma}_{i,i}\right\}\\
	&\geq
	\dfrac{(1-P_{C i,i}-k_0\gamma)^2}{\widehat{\Sigma}_{i,i}} \quad(\mbox{take $L=2(1-P_{C i,i}-k_0\gamma)/\widehat{\Sigma}_{i,i}$}).
	\end{align*}
$\square$

\end{document}